\documentclass[11pt]{article}
\usepackage{typearea}\typearea{14}
\usepackage{epsfig,amsmath,amsfonts,amssymb, cite}
\usepackage[dvipdfmx]{hyperref}%\usepackage{pxjahyper}%these two come together
\hypersetup{hidelinks}
\newtheorem{theorem}{Theorem}[section]
\newtheorem{lemma}[theorem]{Lemma}

\makeatletter
\@addtoreset{equation}{section}
\makeatother

\makeatletter
\long\def\@makecaption#1#2{{\small
\advance\leftskip1cm
\advance\rightskip1cm
\vskip\abovecaptionskip
\sbox\@tempboxa{#1: #2}%
\ifdim \wd\@tempboxa >\hsize
 #1: #2\par
\else
\global \@minipagefalse
\hb@xt@\hsize{\hfil\box\@tempboxa\hfil}%
\fi
\vskip\belowcaptionskip}}
\makeatother
\def\eq#1\en{\begin{equation}#1\end{equation}}  
\def\eqa#1\ena{\begin{align}#1\end{align}}
\def\eqg#1\eng{\begin{gather}#1\end{gather}}
\newcommand{\lb}[1]{\label{e:#1}}
\newcommand{\rlb}[1]{\eqref{e:#1}} 
\newcommand{\nl}{\notag\\}

\newcommand{\snorm}[1]{\Vert#1\Vert}

\newcommand{\sbkt}[1]{\langle#1\rangle}

\newcommand{\sumtwo}[2]%
{\mathop{\sum_{#1}}_{#2}}
\newcommand{\sumthree}[3]%
{\mathop{\mathop{\sum_{#1}}_{#2}}_{#3}}
\newcommand{\sumfour}[4]%
{\mathop{\mathop{\mathop{\sum_{#1}}_{#2}}_{#3}}_{#4}} 
\newcommand{\prodtwo}[2]%
{\mathop{\prod_{#1}}_{#2}}
\newcommand{\mintwo}[2]%
{\mathop{\min_{#1}}_{#2}}
\newcommand{\maxtwo}[2]%
{\mathop{\max_{#1}}_{#2}}
\newcommand{\maxthree}[3]%
{\mathop{\mathop{\max_{#1}}_{#2}}_{#3}}
\newcommand{\limtwo}[2]%
{\mathop{\lim_{#1}}_{#2}}
\newcommand{\suptwo}[2]%
{\mathop{\sup_{#1}}_{#2}}
\newcommand{\supthree}[3]%
{\mathop{\mathop{\sup_{#1}}_{#2}}_{#3}}
\newcommand{\supfour}[4]%
{\mathop{\mathop{\mathop{\sup_{#1}}_{#2}}_{#3}}_{#4}} 
\newcommand{\inftwo}[2]%
{\mathop{\inf_{#1}}_{#2}}
\newcommand{\infthree}[3]%
{\mathop{\mathop{\inf_{#1}}_{#2}}_{#3}}
\newcommand{\inffour}[4]%
{\mathop{\mathop{\mathop{\inf_{#1}}_{#2}}_{#3}}_{#4}} 
\newcommand\calA{{\cal A}}

\newcommand\calH{{\cal H}}

\newcommand\calS{{\cal S}}

\newcommand{\bss}{\boldsymbol{\sigma}}

\newcommand{\bbC}{\mathbb{C}}

\newcommand{\bbZ}{\mathbb{Z}}
\newcommand{\ep}{\varepsilon}
\newcommand{\up}{\uparrow}

\newcommand{\ket}[1]{|#1\rangle}
\newcommand{\bra}[1]{\langle#1|}
\newcommand{\La}{\Lambda}
\newcommand{\Htot}{\calH_\mathrm{tot}}
\newcommand{\Dtot}{D_\mathrm{tot}}

\newcommand{\hP}{\hat{P}}
\newcommand{\Pneq}{\hat{P}_\mathrm{neq}}
\newcommand{\hA}{\hat{A}}
\newcommand{\hH}{\hat{H}}

\newcommand{\hV}{\hat{V}}

\newcommand{\kPz}{\ket{\Phi(0)}}
\newcommand{\bPz}{\bra{\Phi(0)}}
\newcommand{\kPt}{\ket{\Phi(t)}}
\newcommand{\bPt}{\bra{\Phi(t)}}
\newcommand{\Tr}{\operatorname{Tr}}

\newcommand{\LaL}{\Lambda_L}
\newcommand{\hS}{\hat{S}}
\newcommand{\hSz}{\hS^{(\mathrm{z})}}

\newcommand{\uket}{\ket{+}}
\newcommand{\dket}{\ket{-}}

\newcommand{\msb}{m_\mathrm{s}(\beta)}
\newcommand{\Es}{E_{\bss}}
\newcommand{\la}{\lambda}

\newcommand{\SL}{\calS_L}

\newcommand{\sH}{\operatorname{spec}(\hH_L)}

\newcommand{\tcalH}{\widetilde{\calH}}

\begin{document}

\noindent
{\large\bf
Macroscopic thermalization by unitary time evolution in the weakly perturbed two-dimensional Ising model --- An application of the Roos-Sugimoto-Teufel-Tumulka-Vogel theory}

\renewcommand{\thefootnote}{\fnsymbol{footnote}}
\medskip\noindent
Hal Tasaki\footnote{%
Department of Physics, Gakushuin University, Mejiro, Toshima-ku,
Tokyo 171-8588, Japan, \\\tt{hal.tasaki@gmail.com}
}
\renewcommand{\thefootnote}{\arabic{footnote}}
\setcounter{footnote}{0}

\begin{quotation}
\small
By applying the theory recently developed by Roos, Sugimoto, Teufel, Tumulka, and Vogel \cite{RoosSugimotoTeufelTumulkaVogel2025}, we construct a rigorous example of an isolated macroscopic quantum system in which every initial state chosen from a microcanonical energy shell exhibits thermalization solely by the unitary time evolution determined by the Hamiltonian.

We consider the Hamiltonian $\hH_L$ of the standard two-dimensional ferromagnetic Ising model with the plus boundary conditions and perturb it with a small self-adjoint operator $\la_L\hV_L$ drawn randomly from the space of self-adjoint operators on the whole Hilbert space.
The Roos-Sugimoto-Teufel-Tumulka-Vogel theory then guarantees that, for all but an exponentially small fraction of the random perturbations, the model exhibits strong ETH (energy eigenstate thermalization hypothesis), i.e., every energy eigenstate in a microcanonical energy shell is in thermal equilibrium in the sense of large deviation, as precisely stated in the main text.

It is then standard that the strong ETH implies the following result for thermalization.
Suppose that the initial state $\kPz$ is chosen from the microcanonical energy shell and may be very far from thermal equilibrium.
Then the time-evolved state $\kPt=e^{-i(\hH_L+\la_L\hV_L)t}\kPz$ is in thermal equilibrium at sufficiently long and typical times $t$, in the sense that the measurement result of the magnetization density almost certainly coincides with the corresponding spontaneous magnetization.

When we state ``typical'' and ``almost certainly'' above, we mean that the corresponding exceptional fractions are exponentially small in the system size $L$. 
\end{quotation}

%%%%%%%%%%%%%%%%%%%%%%
\section{Introduction}
It is widely accepted by now that an isolated macroscopic quantum system evolving under unitary time evolution can exhibit equilibration or thermalization, i.e., the approach to equilibrium from a nonequilibrium initial state.\footnote{
In the present paper, we understand that ``equilibration'' is a broad notion that stands for approach to equilibrium in general.
When the target is thermal equilibrium, we use the more restricted terminology, ``thermalization''.
Note that the terminology is not unified in the literature (including that of the present author).
\label{fn:TE}
}
See, e.g., \cite{RoosSugimotoTeufelTumulkaVogel2025,vonNeumann,GLTZ,GLMTZ09b,Reimann2015,Tasaki2016,DAlessioKafriPolkovnikovRigol2016,GE16} and references therein.
Recently, there have been ``constructive'' approaches in which one studies a concrete, tractable quantum many-body system and establishes the presence of equilibration (in certain specified senses) without relying on any unproven assumptions \cite{GluzaEisertFarrelly2019,ShiraishiTasaki2023,TasakiFreeFermion,TasakiHeat,RoosSugimotoTeufelTumulkaVogel2025,HaraKoike}.
The present work adds another such example, which may indeed be regarded as an instance of thermalization rather than mere equilibration.

Here we follow the macroscopic approach to thermalization, which originated with von Neumann in \cite{vonNeumann,GLTZ} and was further developed in \cite{GLMTZ09b,Reimann2015,Tasaki2016}.
In this approach, the notion of ETH (energy eigenstate thermalization hypothesis) \cite{vonNeumann, GLTZ,Deutsch1991,Srednicki1994,Tasaki1998,RigolSrednicki2012}, which states that every energy eigenstate in a specified microcanonical energy shell is itself in (thermal) equilibrium, plays a central role.

Recently, Roos, Sugimoto, Teufel, Tumulka, and Vogel developed a general theory of macroscopic equilibration in systems with highly degenerate energy eigenvalues \cite{RoosSugimotoTeufelTumulkaVogel2025}.
Suppose that one can find at least one energy eigenbasis for the unperturbed Hamiltonian in which every basis state is in equilibrium in a certain specified sense.
Note that this assumption differs from ETH, since there may be (and indeed are) energy eigenstates that are far from equilibrium (because of the high degeneracy).
They proved that, for most sufficiently weak generic random perturbations, the perturbed Hamiltonian satisfies ETH.
Combined with their general dynamical result, which derives equilibration from ETH, this shows that every initial state equilibrates for a single typical perturbation.

Roos, Sugimoto, Teufel, Tumulka, and Vogel applied their general theorem to weakly perturbed free fermions in two or higher dimensions and proved thermalization for every initial state \cite{RoosSugimotoTeufelTumulkaVogel2025}.
Their theory, however, applies to a much more general class of Hamiltonians with large degeneracies.
We believe that it can be used to discuss various physically nontrivial examples of equilibration or thermalization in isolated macroscopic quantum systems.
Here we present one such example.

We demonstrate the implications of the Roos-Sugimoto-Teufel-Tumulka-Vogel theory by applying it to a simple, nontrivial, and well-known problem, namely, the two-dimensional Ising model.
We focus on the low-temperature phase, where the model exhibits nontrivial spontaneous magnetization.

To be precise, we consider the standard Ising model Hamiltonian $\hH_L$ on the $L\times L$ square lattice with the plus boundary conditions.
We fix any sufficiently large inverse temperature $\beta>0$ and let $E_L(\beta)$ denote the corresponding energy.
We then add a weak perturbation $\la_L\hV_L$ drawn randomly from the whole space of self-adjoint operators.
The following statement is then valid with probability not less than $1-e^{-C(\beta,\delta)L}$, where $C(\beta,\delta)>0$ is a constant that depends only on the inverse temperature $\beta$ and a fixed precision $\delta$.
An arbitrary initial state $\kPz$ chosen from the microcanonical energy shell around the energy $E_L(\beta)$ is driven to thermal equilibrium at sufficiently long and typical times $t$.
More precisely, there exist a sufficiently large $T$ (about which we do not have any estimates) and a set of atypical times $\calA$ whose total length is at most $T e^{-C(\beta,\delta)L}$, such that the measurement result of the magnetization in the time-evolved state $\kPt=e^{-i(\hH_L+\la_L\hV_L)t}\kPz$ coincides (within the specified precision $\delta$) with the spontaneous magnetization $\msb$ with probability not less than $1-e^{-C(\beta,\delta)L}$ whenever $t\in[0,T]\backslash\calA$.

Apart from the Roos-Sugimoto-Teufel-Tumulka-Vogel theory, our proof relies on the surface-order large-deviation upper bound for the magnetization in the two-dimensional Ising model by Pfister \cite{Pfister1991} and Pfister and Velenik \cite{PfisterVelenik1997}, as well as general considerations on thermalization in \cite{Tasaki2016} based on the large-deviation point of view.

We should stress that our perturbation, which is drawn randomly from the whole space of self-adjoint operators of the system, generally contains highly nonlocal many-spin interactions, including terms of macroscopic range and terms involving a macroscopic number of spins.
We nevertheless expect that it captures some aspects of generic perturbations that make the model non-integrable.
In our theorem, the operator norm of the entire perturbation is bounded uniformly in the system size.
In this sense, our random perturbation is much smaller than the random fields normally considered in random magnetic systems, in which the local field strengths are kept fixed and the norm of the whole random-field term is typically of order $L^2$.

The present paper is organized as follows.
In Section~2, we introduce the Ising model, formulate the thermodynamic large-deviation bound, and state the main thermalization theorem.
In Section~3, we prove the large-deviation bound and the main theorem by applying the Roos-Sugimoto-Teufel-Tumulka-Vogel theory.
Section~4 is devoted to discussion.

%%%%%%%%%%%%%%%%%%%%%%
%%%%%%%%%%%%%%%%%%%%%%
\section{Setting and the main result}
\subsection{The two-dimensional Ising model}

We consider a quantum spin system in which a spin $S=1/2$ is associated with every site of the $L\times L$ square lattice
\eq
\LaL=\{1,\ldots,L\}^2\subset\bbZ^2.
\en
The Hilbert space of the whole system is $\Htot=\bigotimes_{x\in\LaL}\mathfrak{h}_x$, where the local Hilbert space is $\mathfrak{h}_x\cong\bbC^2$.
Let $\hat{\boldsymbol{S}}_x=(\hS^{(\mathrm{x})}_x,\hS^{(\mathrm{y})}_x,\hS^{(\mathrm{z})}_x)$ denote the spin operator at site $x$.
The elements of the standard orthonormal basis $\{\uket_x,\dket_x\}$ of $\mathfrak{h}_x$ satisfy
\eq
\hSz_x\uket_x=\frac{1}{2}\uket_x,\quad
\hSz_x\dket_x=-\frac{1}{2}\dket_x.
\en

A collection $\bss=(\sigma_x)_{x\in\LaL}$ of spin variables $\sigma_x\in\{+,-\}$ is called a classical spin configuration.
We denote by $\SL=\{+,-\}^{\LaL}$ the set of all spin configurations on $\LaL$.
For $\bss\in\SL$, we define the corresponding spin state by
\eq
\ket{\bss}=\bigotimes_{x\in\LaL}\ket{\sigma_x}_x.
\en
The collection $\{\ket{\bss}\}_{\bss\in\SL}$ forms a basis of $\Htot$.
The dimension of $\Htot$ is $\Dtot=2^{L^2}$.

\begin{figure}
\centerline{\epsfig{file=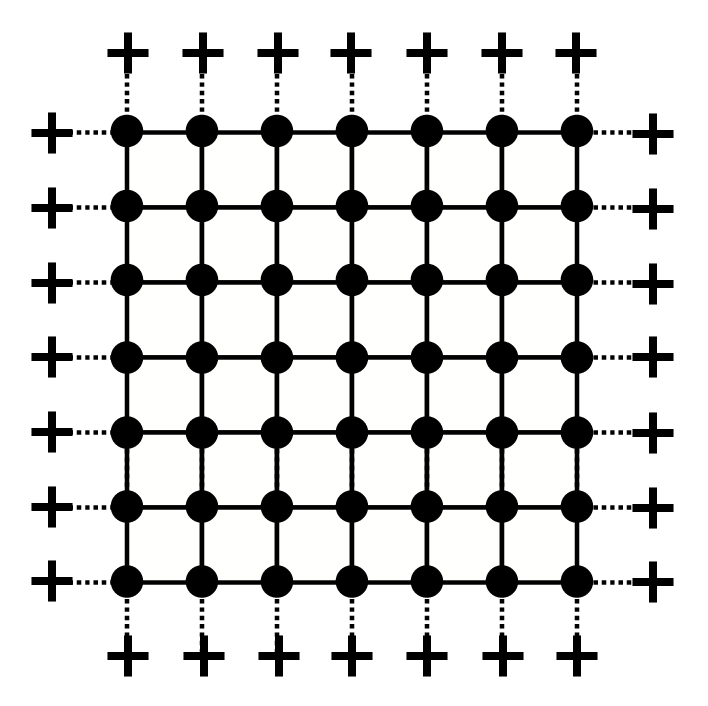,width=5.5truecm}}
\caption[dummy]{
Black dots represent sites in the $7\times7$ square lattice $\La_7$.
We suppose that the lattice is surrounded by fixed $+$ spins, whose effect is represented in the second summation in \rlb{HL}.
The quantity $n_x$ represents the number of dashed lines attached to the site $x$.
We have $n_x=2$ for the four sites at the corners, $n_x=1$ for the twenty sites at the edges (but not on the corners), and $n_x=0$ for the remaining twenty-five interior sites.
}
\label{f:La7}
\end{figure}

The Hamiltonian of the standard Ising model with nearest-neighbor ferromagnetic interactions and the plus boundary conditions\footnote{
We use the plus boundary conditions to select the symmetry-broken plus phase in the infinite-volume limit.
At sufficiently high temperatures, where the Gibbs state is unique, one may instead use periodic boundary conditions; see Section~\ref{s:discussion}.
} is
\eq
\hH_L=-\sumtwo{\{x,y\}\subset\LaL}{(|x-y|=1)}\hSz_x\hSz_y-\sum_{x\in\LaL}\frac{n_x}{2}\hSz_x.
\lb{HL}
\en
Note that a neighboring pair of sites, $x$ and $y$, is counted only once in the first summation.
The second summation represents the effect of the plus boundary conditions.
Here, $n_x$ denotes the number of $u\in\bbZ^2\backslash\LaL$ such that $|x-u|=1$, i.e., the number of neighboring sites to $x$ outside $\LaL$.
See Figure~\ref{f:La7}.

The canonical expectation of the model at inverse temperature $\beta>0$ is
\eq
\sbkt{\,\cdot\,}_{L,\beta}^\mathrm{can}=\frac{\Tr[\ \cdot\ e^{-\beta\hH_L}]}{Z_L(\beta)},
\en
where the partition function is
\eq
Z_L(\beta)=\Tr[e^{-\beta\hH_L}].
\en
It is well known that the model undergoes a phase transition characterized by the spontaneous magnetization $\msb$ as
\eq
\lim_{L\up\infty}\Bigl\langle\frac{1}{L^2}\sum_{x\in\LaL}\hSz_x\Bigr\rangle_{L,\beta}^\mathrm{can}=
\begin{cases}
0,&\beta\le\beta_\mathrm{c};\\
\msb>0,&\beta>\beta_\mathrm{c},
\end{cases}
\en
where the critical inverse temperature is $\beta_\mathrm{c}=2\log(\sqrt{2}+1)$.
Here, we shall focus only on the low-temperature region, where $\msb$ is nonzero.

\subsection{Energy eigenstates and the microcanonical expectation}
For any classical spin configuration $\bss\in\SL$, the state $\ket{\bss}$ satisfies
\eq
\hH_L\ket{\bss}=\Es\ket{\bss},
\lb{EVE}
\en
with the corresponding energy eigenvalue\footnote{%
We here slightly abuse notation and regard $\sigma_x\in\{1,-1\}$.
}
\eq
\Es=-\frac{1}{4}\sumtwo{\{x,y\}\subset\LaL}{(|x-y|=1)}\sigma_x\sigma_y-\frac{1}{4}\sum_{x\in\LaL}n_x\sigma_x.
\en
It is easily found that $\Es\in\bbZ$ and\footnote{%
Since there are $2L(L+1)$ bonds (including the dashed lines in Figure~\ref{f:La7}), the ground-state energy is $-L(L+1)/2$.
In an excited state, each ``unhappy'' bond costs extra energy $1/2$.
Since the number of unhappy bonds is even, $E_{\bss}$ is always an integer.
}
\eq
-\frac{L(L+1)}{2}\le\Es<\frac{L(L+1)}{2}.
\lb{Esbound}
\en
Let us denote by $\sH\subset\bbZ$ the set of all energy eigenvalues of $\hH_L$.

Clearly, the eigenvalues of $\hH_L$ are highly degenerate.
For $E\in\sH$, we denote by $D_L(E)$ its degeneracy.
The high degeneracy makes it natural to define the microcanonical expectation for this model as
\eq
\sbkt{\,\cdot\,}_{L,E}^\mathrm{mc}=\frac{1}{D_L(E)}\sumtwo{\bss\in\SL}{(\Es=E)}\bra{\bss}\cdot\ket{\bss},
\en
for any $E\in\sH$.
Note that we do not introduce a nonzero width for the energy interval.

For any $\beta>0$ and $L$, we define $E_L(\beta)$ as an energy $E\in\sH$ that maximizes $D_L(E)e^{-\beta E}$.
According to the equivalence of ensembles, this $E_L(\beta)$ is the energy of the microcanonical ensemble that corresponds to the canonical ensemble at inverse temperature $\beta$.
See, e.g., \cite{TasakiEQE} and references therein.

%%%%%%%%%%%%%%%%%%%%%%
\subsection{Characterization of thermal equilibrium and the large-deviation bound}
\label{s:LD}
Let us discuss how we characterize thermal equilibrium in the model.
Fix an arbitrary $\beta>\beta_\mathrm{c}$.
We take an initial state whose energy is $E_L(\beta)$ (or close to $E_L(\beta)$ after the perturbation) and let it evolve according to the unitary time evolution.
We expect the system to reach thermal equilibrium at inverse temperature $\beta$.
We also assume, for simplicity, that one is interested only in a single (but most interesting) macroscopic observable, namely, the magnetization density $L^{-2}\sum_{x\in\LaL}\hSz_x$.
It is straightforward to extend the theory to cover finitely many, or a suitably controlled $L$-dependent number of, macroscopic observables.
This is why we call the present result ``thermalization'' rather than ``equilibration''.
See footnote~\ref{fn:TE} and Section~\ref{s:discussion}.

Given a normalized pure state $\ket{\Phi}\in\Htot$, one makes a projective measurement of the magnetization density.
If the measurement result coincides with the expected equilibrium value $\msb$ within a small fixed precision $\delta\in(0,\msb)$, we conclude that the state is in thermal equilibrium.
This motivates us to define the nonequilibrium projection operator as
\eq
\Pneq=\operatorname{Proj}\biggl[\Bigl|\frac{1}{L^2}\sum_{x\in\LaL}\hSz_x-\msb\Bigr|>\delta\biggr].
\lb{Pneq}
\en
Then, by definition, $\bra{\Phi}\Pneq\ket{\Phi}$ is the probability that the state $\ket{\Phi}$ is not found in thermal equilibrium at $\beta$.
By slightly extending the use of the terminology, we say that a normalized pure state $\ket{\Phi}$ (with energy close to $E_L(\beta)$) is in thermal equilibrium\footnote{%
This is called MATE (macroscopic thermal equilibrium) in \cite{RoosSugimotoTeufelTumulkaVogel2025}.
} if $\bra{\Phi}\Pneq\ket{\Phi}$ is negligibly small.
As we shall discuss in Section~\ref{s:discussion}, the notion of thermal equilibrium can be strengthened by redefining the projection operator $\Pneq$.

The following essential lemma establishes a large-deviation upper bound for the microcanonical expectation value of $\Pneq$.
This type of inequality is called the thermodynamic bound in \cite{Tasaki2016}.\footnote{
Note that here the exponential scale is proportional to $L$, rather than to $L^2$ (where only the latter scale is discussed in  \cite{Tasaki2016}).
Indeed, a fixed macroscopic deficit of the plus-phase magnetization can be realized by a region of the minus phase whose area is of order $L^2$ but whose boundary has length of order $L$.
The corresponding free-energy cost is the surface tension times the interface length and is therefore of order $L$.}
As we shall see in Section~\ref{s:proof1}, the lemma is an immediate consequence of phase-separation large-deviation results of Pfister \cite{Pfister1991} and of Pfister and Velenik \cite{PfisterVelenik1997}.

\begin{lemma}\label{LD}
For sufficiently large $\beta$, there are constants $C(\beta,\delta)>0$ and $L_0(\beta,\delta)$ such that
\eq
\sbkt{\Pneq}_{L,E_L(\beta)}^\mathrm{mc}\le e^{-8C(\beta,\delta)L}
\lb{mcLD}
\en
for any $L\ge L_0(\beta,\delta)$.
\end{lemma}

Here, and in what follows, we assume $L$ is large enough that $C(\beta,\delta)L\gg1$ and the right-hand side of \rlb{mcLD} is negligibly small.
Then the above lemma states that the distribution of the magnetization density $L^{-2}\sum_{x\in\LaL}\hSz_x$ is sharply concentrated around $\msb$ in the microcanonical ensemble for energy $E_L(\beta)$.
This is a manifestation of the equivalence of the microcanonical ensemble with energy $E_L(\beta)$ and the canonical ensemble with inverse temperature $\beta$.

We recall that the thermodynamic bound \rlb{mcLD} has a direct implication on the typicality of thermal equilibrium \cite{Tasaki2016}.
Let $\calH_{L,\beta}$ be the (highly degenerate) eigenspace of $\hH_L$ corresponding to energy $E=E_L(\beta)$.
Suppose one uniformly draws a normalized state $\ket{\Phi}$ from $\calH_{L,\beta}$.
Then it is an immediate consequence of \rlb{mcLD} that, with probability not less than $1-e^{-7C(\beta,\delta)L}$, the randomly chosen state $\ket{\Phi}$ satisfies
\eq
\bra{\Phi}\Pneq\ket{\Phi}\le e^{-C(\beta,\delta)L}.
\en
This means that the pure state $\ket{\Phi}$ is in thermal equilibrium in the sense discussed above.

%%%%%%%%%%%%%%%%%%%%%%
\subsection{Unitary time evolution and thermalization under weak random perturbation}
\label{s:U}
The Ising Hamiltonian \rlb{HL}, which is essentially classical, generates only trivial dynamics in the classical spin basis.
Indeed, it follows from \rlb{EVE} that each state $\ket{\bss}$ is stationary up to an overall phase under the time evolution determined by $\hH_L$.
To generate nontrivial dynamics, we perturb $\hH_L$ by a small ``quantum'' interaction $\la_L\hV_L$, where $\hV_L$ is a self-adjoint operator and $\la_L>0$ is a small parameter.
For a normalized initial state $\kPz$, the state at time $t$ is given by
\eq
\kPt=e^{-i(\hH_L+\la_L\hV_L)t}\kPz.
\lb{Pt}
\en
We expect that, for a suitable choice of $\hV_L$, the time-evolved state $\kPt$ is in thermal equilibrium for sufficiently large and typical $t$.

We conjecture that any $\hV_L$ that makes the total Hamiltonian $\hH_L+\la_L\hV_L$ non-integrable will lead to thermalization.
Realistic examples include the transverse magnetic field $\hV_L=-\sum_{x\in\LaL}\hS^{\mathrm{x}}_x$ or the XY interaction $\hV_L=-\sum_{\{x,y\}\subset\LaL\,(|x-y|=1)}\{\hS^{\mathrm{x}}_x\hS^{\mathrm{x}}_y+\hS^{\mathrm{y}}_x\hS^{\mathrm{y}}_y\}$, but to treat these models rigorously seems to be formidably difficult.\footnote{
Rigorous results in \cite{Chiba2025} and \cite{ShiraishiTasaki2026} strongly suggest that these models are non-integrable.
}
Roos, Sugimoto, Teufel, Tumulka, and Vogel proposed considering a generic random perturbation and solved the corresponding problem \cite{RoosSugimotoTeufelTumulkaVogel2025}.

Following \cite{RoosSugimotoTeufelTumulkaVogel2025}, we regard the self-adjoint operators on $\Htot$ as a finite-dimensional real vector space equipped with Lebesgue measure, and draw $\hV_L$ from the normalized Lebesgue measure on the operator-norm unit ball
\eq
\{\hV\,|\, \hV=\hV^\dagger,\,\snorm{\hV}\le1\}.
\lb{Vball}
\en
This is a continuous probability distribution invariant under conjugation by every unitary operator on $\Htot$.
Note that such a random $\hV_L$ contains, with probability one, highly nonlocal many-spin interactions, including terms of macroscopic range and terms involving a macroscopic number of spins.
It can hardly be regarded as a realistic perturbation.
We nevertheless expect that a random $\hV_L$ may represent certain generic features, including thermalization, exhibited by realistic short-range perturbations.
See Section~2.4 of \cite{RoosSugimotoTeufelTumulkaVogel2025}.

Let us define the microcanonical energy shell of the perturbed Hamiltonian as
\eq
\tcalH_{L,\beta}=\operatorname{span}\Bigl\{\ket{\psi}\,\Bigl|\,(\hH_L+\la_L\hV_L)\ket{\psi}=E\ket{\psi},\,E\in(E_L(\beta)-\tfrac12,E_L(\beta)+\tfrac12)\Bigr\}.
\lb{HLb}
\en
Note that, for $\la_L=0$, the energy shell $\tcalH_{L,\beta}$ reduces to the energy eigenspace $\calH_{L,\beta}$ defined above.
In the definition \rlb{HLb}, we anticipated that the perturbation $\la_L\hV_L$ is small enough that it essentially lifts the degeneracy of the Ising Hamiltonian $\hH_L$ (and slightly mixes the different eigenspaces of $\hH_L$).

The following theorem is a consequence of the argument behind Theorem~1 and Proposition~1 of \cite{RoosSugimotoTeufelTumulkaVogel2025}.

\begin{theorem}\label{main}
Take sufficiently large $\beta$ for which Lemma~\ref{LD} is valid and fix $\delta\in(0,\msb)$.
Let $C(\beta,\delta)>0$ be the constant in Lemma~\ref{LD}.
For all sufficiently large $L$, one can choose $0<\la_L<1/4$ such that, with probability not less than $1-e^{-C(\beta,\delta)L}$ for the random choice of $\hV_L$, the following statement is valid for every normalized initial state $\kPz\in\tcalH_{L,\beta}$.
For each such initial state, there exist a sufficiently large $T$ and a subset $\calA\subset[0,T]$ such that
\eq
\frac{\mu(\calA)}{T}\le e^{-C(\beta,\delta)L},
\lb{muA}
\en
where $\mu(\calA)$ is the Lebesgue measure (the total length) of $\calA$, and one has
\eq
\bPt\Pneq\kPt\le e^{-C(\beta,\delta)L},
\lb{PtEQ}
\en
for any $t\in[0,T]\backslash\calA$, where $\kPt$ is defined by \rlb{Pt}.
\end{theorem}

Recall that we assume $C(\beta,\delta) L\gg1$, and hence $e^{-C(\beta,\delta)L}$ is negligibly small.
The event in Theorem~\ref{main} has probability close to one for large $L$.
Fix a perturbation $\hV_L$ in this event.
It follows from \rlb{muA} that $\calA$ occupies only an exponentially small fraction of the interval $[0,T]$.
Thus a measurement time chosen uniformly from $[0,T]$ lies outside $\calA$ with probability at least $1-e^{-C(\beta,\delta)L}$.
For any such time, \rlb{PtEQ} implies that a projective measurement of the magnetization density $L^{-2}\sum_{x\in\LaL}\hSz_x$ in the pure state $\kPt$ agrees with the spontaneous magnetization $\msb$ within the precision $\delta>0$, with probability not less than $1-e^{-C(\beta,\delta)L}$.
Thus the pure state $\kPt$ is in thermal equilibrium in the sense discussed in Section~\ref{s:LD}.

To sum up, the theorem establishes that the state $\kPt$ is found in thermal equilibrium at a sufficiently long and typical time $t$.
We stress that the initial state $\kPz\in\tcalH_{L,\beta}$ can be very far from equilibrium in the sense that the magnetization density $L^{-2}\sum_{x\in\LaL}\hSz_x$ is drastically different from the equilibrium value $\msb$.
One can even take $\kPz$ to be a Schr\"odinger-cat state, i.e., a superposition of states with macroscopically distinct magnetizations.

Unfortunately, our theorem, as well as other rigorous equilibration results based on ETH, does not provide any information on the time scale $T$ of thermalization.\footnote{
In the present model, the perturbation is highly nonlocal and the proof gives no quantitative lower bound on its admissible strength, so such a time scale would in any case have limited physical significance.
}
This is an essential limitation of the current approach.
See \cite{GluzaEisertFarrelly2019,HaraKoike} for treatments of the time scale in free fermion systems.

\medskip\noindent
{\em Remarks:}\/
1.~Let us also mention the degeneracy of the perturbed Hamiltonian.
For every fixed $L$ and every fixed $\la_L>0$, the distribution of $\hH_L+\la_L\hV_L$ is absolutely continuous in the space of self-adjoint operators.
It follows that its energy eigenvalues are nondegenerate with probability one; see Lemma~2 of \cite{RoosSugimotoTeufelTumulkaVogel2025}.
We do not use this fact in the proof of Theorem~\ref{main}.

2.~In Theorem~\ref{main}, we assumed that the initial state $\kPz$ is chosen from the microcanonical energy shell $\tcalH_{L,\beta}$ of the perturbed Hamiltonian.
This is the standard formulation from the viewpoint of the foundations of statistical mechanics.
From the viewpoint of state preparation, however, it may be more natural to choose the initial state from the eigenspace $\calH_{L,\beta}$ of the unperturbed Ising Hamiltonian; one may then take, for example, a classical spin state $\ket{\bss}$ with $E_{\bss}=E_L(\beta)$.
A corresponding version of Theorem~\ref{main} can also be proved.
Indeed, since the unperturbed level $E_L(\beta)$ is separated from all other levels by a gap of at least one and $\snorm{\hV_L}\le1$, the spectral projections onto $\tcalH_{L,\beta}$ and $\calH_{L,\beta}$ differ by $O(\la_L)$, uniformly in $\hV_L$.
The perturbative conclusion remains valid if $\la_L$ is decreased further.
By imposing an additional $L$-dependent upper bound on $\la_L$, the difference between the two energy shells can therefore be absorbed into the error terms, yielding the analogue of Theorem~\ref{main} with $\tcalH_{L,\beta}$ replaced by $\calH_{L,\beta}$.
We do not pursue this variant here.

%%%%%%%%%%%%%%%%%%%%%%
%%%%%%%%%%%%%%%%%%%%%%
\section{Proofs}
\subsection{Proof of Lemma~\ref{LD}}\label{s:proof1}
The nontrivial thermodynamic input is the following surface-order large-deviation bound. It follows from the phase-separation theory of Pfister \cite{Pfister1991} and of Pfister and Velenik \cite{PfisterVelenik1997}.

\begin{theorem}\label{P}
For sufficiently large $\beta$, there are constants $\tilde C(\beta,\delta)>0$ and $\tilde L_0(\beta,\delta)$ such that
\eq
\sbkt{\Pneq}_{L,\beta}^\mathrm{can}\le e^{-\tilde C(\beta,\delta) L}
\lb{canLD}
\en
for any $L\ge\tilde L_0(\beta,\delta)$.
\end{theorem}

Let us recall the origin of the scale in \rlb{canLD}.
The lower tail of the magnetization is in the phase-coexistence region.
A fixed deficit from the plus-phase spontaneous magnetization is obtained by forming a macroscopic droplet of the minus phase.
Its area is proportional to $L^2$, while the free-energy cost is governed by the surface tension and is proportional to the length of its boundary, which is of order $L$.
The corresponding large-deviation probability is therefore $e^{-O(L)}$.
The upper tail above $\msb+\delta$ has a positive bulk large-deviation rate and is smaller still.
The phase-separation results in \cite{Pfister1991,PfisterVelenik1997} make this mechanism rigorous and imply Theorem~\ref{P}.

Our only task is to relate the expectation value $\sbkt{\Pneq}_{L,\beta}^\mathrm{can}$ to that in the microcanonical ensemble.
Although the procedure is discussed in Section~8.1 of \cite{Tasaki2016} in the general setting, we shall give a complete derivation here since the argument simplifies considerably in the present setting.

Note first that the partition function is written as
\eq
Z_L(\beta)=\sum_{E\in\sH}D_L(E)e^{-\beta E}.
\en
Recalling that $E=E_L(\beta)$ maximizes $D_L(E)e^{-\beta E}$, we find
\eq
Z_L(\beta)\le L(L+1)D_L(E_L(\beta))e^{-\beta E_L(\beta)},
\en
where we used \rlb{Esbound} to bound the number of elements of $\sH$ by $L(L+1)$.
We then observe
\eqa
\sbkt{\Pneq}_{L,\beta}^\mathrm{can}
&=\frac{1}{Z_L(\beta)}\sum_{\bss\in\SL}e^{-\beta E_{\bss}}\bra{\bss}\Pneq\ket{\bss}
\nl
&\ge\frac{e^{-\beta E_L(\beta)}}{Z_L(\beta)}\sumtwo{\bss\in\SL}{(E_{\bss}=E_L(\beta))}\bra{\bss}\Pneq\ket{\bss}
\nl
&=\frac{D_L(E_L(\beta))e^{-\beta E_L(\beta)}}{Z_L(\beta)}\sbkt{\Pneq}_{L,E_L(\beta)}^\mathrm{mc}
\nl
&\ge\frac{1}{L(L+1)}\sbkt{\Pneq}_{L,E_L(\beta)}^\mathrm{mc}.
\ena
The desired large-deviation bound \rlb{mcLD} follows from Theorem~\ref{P} by decreasing the constant in the exponent and taking $L$ sufficiently large.

%%%%%%%%%%%%%%%
\subsection{Main lemma and the proof of Theorem~\ref{main}}
The main ingredient of the present theory is the following lemma, which follows by applying the argument of Roos, Sugimoto, Teufel, Tumulka, and Vogel \cite{RoosSugimotoTeufelTumulkaVogel2025} to the single energy eigenspace $\calH_{L,\beta}$ corresponding to $E_L(\beta)$.

\begin{lemma}\label{mainL}
Take sufficiently large $\beta$ for which Lemma~\ref{LD} is valid and fix $\delta\in(0,\msb)$.
Let $C(\beta,\delta)>0$ be the constant in Lemma~\ref{LD}.
For all sufficiently large $L$, one can choose $0<\la_L<1/4$ such that, with probability not less than $1-e^{-C(\beta,\delta)L}$ for the random choice of $\hV_L$, every normalized eigenstate $\ket{\Psi}$ of the perturbed Hamiltonian $\hH_L+\la_L\hV_L$ in the microcanonical energy shell $\tcalH_{L,\beta}$ satisfies
\eq
\bra{\Psi}\Pneq\ket{\Psi}\le\frac{1}{2}e^{-2C(\beta,\delta)L}.
\lb{ETH}
\en
\end{lemma}
Note that \rlb{ETH} is ETH in the large-deviation form \cite{Tasaki2016}.

We shall prove the lemma in the next subsection.
Given the lemma, the proof of Theorem~\ref{main} is standard.
Let us discuss the proof for completeness.

Set $\ep_L=e^{-C(\beta,\delta)L}$ and write $\widetilde H_L=\hH_L+\la_L\hV_L$.
Let $\{\widetilde E_a\}_a$ be the distinct eigenvalues of $\widetilde H_L$ in the interval $(E_L(\beta)-1/2,E_L(\beta)+1/2)$, and let $\widetilde P_a$ be the corresponding spectral projections.
Since every normalized vector in $\operatorname{Ran}\widetilde P_a$ is an eigenstate in $\tcalH_{L,\beta}$, Lemma~\ref{mainL} implies
\eq
\snorm{\widetilde P_a\Pneq\widetilde P_a}\le\frac{\ep_L^2}{2}
\lb{blockETH}
\en
for every $a$.
This is the point where the formulation of Proposition~1 of \cite{RoosSugimotoTeufelTumulkaVogel2025}, which allows energy degeneracy, is useful.

Take an arbitrary normalized initial state $\kPz\in\tcalH_{L,\beta}$.
Since $\tcalH_{L,\beta}$ is invariant under $\widetilde H_L$, the standard time-average calculation gives
\eqa
\lim_{T\up\infty}\frac{1}{T}\int_0^T\bPt\Pneq\kPt\,dt
&=\sum_a\bPz\widetilde P_a\Pneq\widetilde P_a\kPz
\nl
&\le\frac{\ep_L^2}{2}\sum_a\bPz\widetilde P_a\kPz
=\frac{\ep_L^2}{2},
\lb{timeaverage}
\ena
where we used \rlb{blockETH}.
Consequently, one can take a sufficiently large $T$ such that
\eq
\frac{1}{T}\int_0^T\bPt\Pneq\kPt\,dt\le\ep_L^2.
\lb{finiteaverage}
\en
Define
\eq
\calA=\{t\in[0,T]~|~\bPt\Pneq\kPt>\ep_L\}.
\en
Then \rlb{finiteaverage} implies
\eq
\frac{\mu(\calA)}{T}\,\ep_L
\le\frac{1}{T}\int_0^T\bPt\Pneq\kPt\,dt
\le\ep_L^2,
\en
and hence $\mu(\calA)/T\le\ep_L$.
For every $t\in[0,T]\backslash\calA$, one has $\bPt\Pneq\kPt\le\ep_L$ by definition.
These are precisely \rlb{muA} and \rlb{PtEQ}.
Notice that the good set of perturbations is independent of the initial state, whereas $T$ and $\calA$ may depend on it.

%%%%%%%%%%%%%%%
\subsection{Proof of Lemma~\ref{mainL}}
Recall that $\calH_{L,\beta}$ is the eigenspace of the unperturbed Hamiltonian $\hH_L$ with eigenvalue $E_L(\beta)$, and set
\eq
D_0=D_L(E_L(\beta))=\dim\calH_{L,\beta}.
\en
The natural basis of $\calH_{L,\beta}$ is $\{\ket{\bss}\,|\,E_{\bss}=E_L(\beta)\}$, but it need not have the desired equilibrium property.
Let us denote the corresponding spin configurations, in an arbitrary order, by $\bss_1,\ldots,\bss_{D_0}$, and define
\eq
\ket{\Psi_\ell}=\frac{1}{\sqrt{D_0}}\sum_{j=1}^{D_0}\exp\Bigl[i\frac{2\pi j\ell}{D_0}\Bigr]\ket{\bss_j},
\lb{Fourierbasis}
\en
for $\ell=1,\ldots,D_0$.
Obviously, $\{\ket{\Psi_\ell}\}_{\ell=1,\ldots,D_0}$ is an orthonormal basis of $\calH_{L,\beta}$.
Moreover, for any operator $\hA$ that is a function only of $\hSz_x$ with $x\in\LaL$, one has
\eq
\bra{\Psi_\ell}\hA\ket{\Psi_\ell}
=\frac{1}{D_0}\sum_{j=1}^{D_0}\bra{\bss_j}\hA\ket{\bss_j}
=\sbkt{\hA}_{L,E_L(\beta)}^\mathrm{mc}.
\lb{basisMC}
\en
In particular, Lemma~\ref{LD} gives
\eq
\bra{\Psi_\ell}\Pneq\ket{\Psi_\ell}\le\ep_0,
\qquad
\ep_0=e^{-8C(\beta,\delta)L},
\lb{goodbasis}
\en
for every $\ell$.

We now apply the concentration estimate of Roos, Sugimoto, Teufel, Tumulka, and Vogel to this single eigenspace.
Since $\hH_L$ and $\Pneq$ commute, the restriction of $\Pneq$ to $\calH_{L,\beta}$ is again a projection.
Let
\eq
C_\mathrm{R}=\frac{2}{9\pi^3},
\qquad
\eta_L=\frac{1}{8}e^{-2C(\beta,\delta)L}.
\en
For all sufficiently large $L$, one has $C_\mathrm{R}\eta_L^3\ge\ep_0$.
Proposition~4 of \cite{RoosSugimotoTeufelTumulkaVogel2025}, applied to $\calH_{L,\beta}$, then shows that a Haar-random orthonormal basis of this eigenspace satisfies
\eq
\bra{\varphi}\Pneq\ket{\varphi}\le\ep_0+\eta_L
\lb{randomB}
\en
for every one of its basis vectors, except for a set of bases with probability at most
\eqa
b_L
&=2D_0\exp\Bigl[-C_\mathrm{R}\frac{\eta_L^3}{\ep_0}\Bigr]
\nl
&\le 2^{L^2+1}\exp\Bigl[-\frac{C_\mathrm{R}}{512}e^{2C(\beta,\delta)L}\Bigr].
\lb{bL}
\ena
In the second line, we used $D_0\le\Dtot=2^{L^2}$.
In particular, $b_L\le e^{-C(\beta,\delta)L}/2$ for all sufficiently large $L$.

Let $\hP_0$ be the projection onto $\calH_{L,\beta}$.
The random operator $\hP_0\hV_L\hP_0$, regarded as an operator on $\calH_{L,\beta}$, has an absolutely continuous distribution invariant under unitary conjugation.
It therefore has nondegenerate eigenvalues with probability one, and its eigenbasis, modulo phases and ordering, is Haar distributed.
By the standard degenerate perturbation theory, the eigenstates of $\hH_L+\la\hV_L$ with eigenvalues in $(E_L(\beta)-1/2,E_L(\beta)+1/2)$ converge, as $\la\downarrow0$, to this eigenbasis of $\hP_0\hV_L\hP_0$.
This is precisely the perturbative argument used in the proof of Theorem~1 of \cite{RoosSugimotoTeufelTumulkaVogel2025}.
It follows that, for every $0<\alpha<1$, there is $\la_0(L,\alpha)>0$ such that, for all $0<\la<\la_0(L,\alpha)$, with probability not less than $(1-\alpha)(1-b_L)$, every normalized eigenstate $\ket{\Psi}$ in the corresponding perturbed energy shell satisfies
\eq
\bra{\Psi}\Pneq\ket{\Psi}\le\ep_0+2\eta_L.
\lb{perturbedB}
\en

Take $\alpha=e^{-C(\beta,\delta)L}/2$ and choose
\eq
0<\la_L<\min\{\la_0(L,\alpha),\,1/4\}.
\lb{lambdaChoice}
\en
By \rlb{bL}, the probability in question is not less than $1-e^{-C(\beta,\delta)L}$.
The condition $\la_L<1/4$, together with $\snorm{\hV_L}\le1$ and the fact that the unperturbed energies are integers, guarantees that the interval in \rlb{HLb} contains precisely the eigenvalues originating from $E_L(\beta)$.
Finally, for all sufficiently large $L$,
\eq
\ep_0+2\eta_L
=e^{-8C(\beta,\delta)L}+\frac{1}{4}e^{-2C(\beta,\delta)L}
\le\frac{1}{2}e^{-2C(\beta,\delta)L}.
\en
This, together with \rlb{perturbedB}, proves Lemma~\ref{mainL}.

%%%%%%%%%%%%%%%%%%%%%%
%%%%%%%%%%%%%%%%%%%%%%
\section{Discussion}
\label{s:discussion}
We proved that any initial state in the microcanonical energy shell $\tcalH_{L,\beta}$ of the weakly perturbed Ising Hamiltonian thermalizes under the unitary time evolution, in the sense that the magnetization density is close to the corresponding spontaneous magnetization $\msb$ at sufficiently long and typical times.
We stress that this is a remarkably nontrivial phenomenon since the precise value of $\msb$ is determined by the intricate interaction between a macroscopic number of spins.
Moreover, one typical perturbation works for all initial states in the whole energy shell $\tcalH_{L,\beta}$.

The key ingredient in the proof is that the eigenspace of the unperturbed Ising Hamiltonian with energy $E_L(\beta)$ admits the basis $\{\ket{\Psi_\ell}\}$ in \rlb{Fourierbasis}, for which the expectation of every observable that is a function of the $\hSz_x$ coincides exactly with its microcanonical expectation value, as in \rlb{basisMC}.
The concentration estimate and the perturbative argument of Roos, Sugimoto, Teufel, Tumulka, and Vogel then establish ETH, in the form \rlb{ETH}, for every eigenstate in the corresponding energy shell of the perturbed Hamiltonian.
The dynamical conclusion follows from Proposition~1 of \cite{RoosSugimotoTeufelTumulkaVogel2025}, which is valid even in the presence of energy degeneracy.
Although a generic perturbation has nondegenerate energy eigenvalues with probability one, this nondegeneracy is not needed in our argument.

Although magnetization is undoubtedly the most interesting observable in this model, one may, in principle, extend the present theory to cover multiple macroscopic observables.
To be more precise, let $\hA_1,\ldots,\hA_n$ be macroscopic quantities that depend only on $\hSz_x$ ($x\in\LaL$), and denote by $a_1(\beta),\ldots,a_n(\beta)$ the equilibrium values of their densities at inverse temperature $\beta$.
An example of $\hA_j$ (other than the magnetization) is the total energy in a macroscopic subset (say, the left half) of the lattice.
Following Section~2.2 of \cite{Tasaki2016}, define the nonequilibrium projection by
\eq
\Pneq^{(n)}=1-\prod_{j=1}^n\operatorname{Proj}\biggl[\Bigl|\frac{\hA_j}{L^2}-a_j(\beta)\Bigr|\le\delta_j\biggr],
\en
where $\delta_j>0$ is the required precision for $\hA_j/L^2$.
By establishing a large-deviation-type upper bound for the fluctuation of each quantity and using
\eq
\Pneq^{(n)}\le\sum_{j=1}^n\operatorname{Proj}\biggl[\Bigl|\frac{\hA_j}{L^2}-a_j(\beta)\Bigr|>\delta_j\biggr],
\en
we obtain a thermodynamic bound for $\Pneq^{(n)}$ and can repeat the present argument.
It follows that all of the densities $\hA_1/L^2,\ldots,\hA_n/L^2$ are simultaneously close to their equilibrium values at sufficiently long and typical times.
One may also let $n=n_L$ depend on $L$, provided that one can establish a sufficiently strong thermodynamic bound for the resulting projection $\Pneq^{(n_L)}$.
At low temperatures, however, proving the required large-deviation-type bounds may be technically nontrivial.

At sufficiently high temperatures, by contrast, this extension is straightforward.
In the Dobrushin uniqueness regime, Gaussian concentration bounds apply uniformly on finite tori to translation averages of arbitrary bounded local functions; see, e.g., \cite{ChazottesColletRedig2017}.
As an illustration, consider the standard Ising model on the $L\times L$ square lattice with periodic boundary conditions.
Let $g(L)$ be a positive integer-valued function such that $g(L)=o(L)$, and let $S_L$ contain one representative from each translation class of nonempty subsets of $\{1,\ldots,g(L)\}^2$.
Then the number $n_L=|S_L|$ of such classes satisfies $n_L\le 2^{g(L)^2}$, and hence $\log n_L=o(L^2)$.
For $X\in S_L$, define
\eq
\hA_X=\sum_{u\in\LaL}\prod_{v\in X}\hSz_{u+v},
\en
where addition of lattice sites is understood modulo $L$, and let
\eq
 a_{X,\infty}(\beta)=\Bigl\langle\prod_{v\in X}\hSz_v\Bigr\rangle_{\beta,\infty},
\en
where $\sbkt{\,\cdot\,}_{\beta,\infty}$ denotes expectation in the unique infinite-volume Gibbs state.
For every fixed $\delta>0$, the Gaussian concentration bound, together with the exponential convergence of periodic finite-volume expectations to the infinite-volume expectation, gives\footnote{%
We use here the notation of \cite{ChazottesColletRedig2017}.
For $X\in S_L$, set $f_X(\bss)=\prod_{v\in X}(\sigma_v/2)$, where $\sigma_v=\pm1$, and write $S_{\LaL}f_X=\sum_{u\in\LaL}f_X(T_u\bss)$.
Then $\snorm{\underline{\delta}(f_X)}_1=|X|2^{1-|X|}\le1$, and Lemma~7.1 of \cite{ChazottesColletRedig2017} gives $\snorm{\underline{\delta}(S_{\LaL}f_X)}_2^2\le L^2$.
The finite-volume version of their Gaussian concentration estimate therefore yields a bound of order $e^{-c(\beta)\delta^2L^2}$ around the periodic finite-volume expectation.
To compare this expectation with $a_{X,\infty}(\beta)$, translate the support of $X$ to the center of the torus and apply the Dobrushin comparison theorem \cite{RebeschiniVanHandel2014}.
It gives, uniformly in $X\in S_L$,
$\bigl|\langle f_X\rangle_{L,\beta}^{\mathrm{per}}-\langle f_X\rangle_{\beta,\infty}\bigr|\le C_1(\beta)e^{-C_2(\beta)(L-g(L))}$.
Since $g(L)=o(L)$, this difference is exponentially small in $L$ and can be absorbed into the fixed precision $\delta$ for all sufficiently large $L$.
}
\eq
\biggl\langle
\operatorname{Proj}\biggl[\Bigl|\frac{\hA_X}{L^2}-a_{X,\infty}(\beta)\Bigr|>\delta\biggr]
\biggr\rangle_{L,\beta}^{\mathrm{per}}
\le 2e^{-c(\beta)\delta^2L^2},
\en
where $c(\beta)>0$ is independent of $X$ and $L$.
Taking the union bound over $X\in S_L$, and using $g(L)=o(L)$, one obtains
\eq
\sbkt{\Pneq^{(n_L)}}_{L,\beta}^{\mathrm{per}}
\le 2^{g(L)^2+1}e^{-c(\beta)\delta^2L^2}
\le e^{-c'L^2}
\en
for some $c'=c'(\beta,\delta)>0$ and all sufficiently large $L$, where $\Pneq^{(n_L)}$ is defined using the observables $\hA_X$, the equilibrium values $a_{X,\infty}(\beta)$, and the common precision $\delta$.
After the same canonical-to-microcanonical conversion used in Section~\ref{s:proof1}, the Roos-Sugimoto-Teufel-Tumulka-Vogel argument shows that, for a typical sufficiently weak perturbation, every initial state in the corresponding perturbed microcanonical energy shell thermalizes, at sufficiently long and typical times, with respect to all of these $n_L$ observables simultaneously.

Let us also note that, in operator norm, the perturbation considered here is much smaller than the random-field perturbations usually studied in random magnetic systems.
For a conventional random-field term $-\sum_x h_x\hSz_x$ with $L$-independent local amplitudes, the operator norm is of order $L^2$.
Here $\snorm{\la_L\hV_L}\le\la_L<1/4$, so the operator norm of the entire perturbation is bounded uniformly in $L$.
Our proof, however, gives no lower bound on the admissible $\la_L$ that is uniform in $L$, and therefore does not show that $\la_L$ can be chosen bounded away from zero as $L\up\infty$.
A generic $\hV_L$ mixes different energy eigenspaces of the unperturbed Ising Hamiltonian at every nonzero $\la_L$, while lifting the large degeneracy within the target energy level.

As we already stressed in Section~\ref{s:U}, we expect that a more realistic short-range perturbation $\hV_L$ will generally lead to thermalization.
It is a challenging problem to develop techniques that would allow us to treat such a perturbation.
See Appendices~A.3 and A.4 of \cite{ShiraishiTasaki2026} for preliminary rigorous results on operator growth and complex-time evolution in non-integrable quantum spin systems.

\bigskip
\noindent{\small
\textbf{Acknowledgments.}
I thank Stefan Teufel and Roderich Tumulka for valuable discussions, Aernout C. D. van Enter for a useful comment on the scale of the perturbation, and Akira Shimizu for ongoing discussions on the characterization of thermal equilibrium.
The present work is supported by JSPS Grants-in-Aid for Scientific Research Nos.~22K03474 and 25K07171.}

\medskip
\noindent{\small
\textbf{Use of generative artificial intelligence.}
The first version of this manuscript, arXiv:2409.09395v1 (2024), was written without artificial-intelligence assistance.
The present substantially revised version was prepared with substantial assistance from ChatGPT (OpenAI, GPT-5.6 Pro; accessed July--August 2026).
ChatGPT was used to analyze the theorem of Roos, Sugimoto, Teufel, Tumulka, and Vogel, identify and formulate the correct surface-order Ising large-deviation input, check the finite-size estimates, and help revise the exposition.
The author checked the source literature and all AI-assisted material and takes full responsibility for the arguments, references, and final content.}

%%%%%%%%%%

\end{document}